\documentclass[10pt, conference, compsocconf]{IEEEtran}
\IEEEoverridecommandlockouts
\usepackage{booktabs,caption,fixltx2e}
\usepackage[flushleft]{threeparttable}

\usepackage{cite}
\usepackage{graphicx}
\usepackage{float}
\usepackage{url}
\begin{document}
\title{A Deep Learning Approach for Privacy Preservation in Assisted Living}


\author{\IEEEauthorblockN{Ismini Psychoula\IEEEauthorrefmark{1},
Erinc Merdivan\IEEEauthorrefmark{2},
Deepika Singh\IEEEauthorrefmark{2} \IEEEauthorrefmark{4},
Liming Chen\IEEEauthorrefmark{1},
Feng Chen\IEEEauthorrefmark{1},\\
Sten Hanke\IEEEauthorrefmark{2},
Johannes Kropf\IEEEauthorrefmark{2},
Andreas Holzinger\IEEEauthorrefmark{4},
Matthieu Geist\IEEEauthorrefmark{3}
}

\IEEEauthorblockA{\IEEEauthorrefmark{1}School of Computer Science and Informatics,
De Montfort University, Leicester, UK \\ 
Email: \{ismini.psychoula, liming.chen\}@dmu.ac.uk}
\IEEEauthorblockA{\IEEEauthorrefmark{2}Center for Health \& Bioresources, AIT Austrian Institute of Technology, Wiener Neustadt, Austria\\ 
Email: \{erinc.merdivan, deepika.singh, sten.hanke, johannes.kropf\}@ait.ac.at}
 
\IEEEauthorblockA{\IEEEauthorrefmark{4}Holzinger Group, Institute for Medical Informatics/Statistics, Medical University Graz, A-8036 Graz, Austria\\
Email: a.holzinger@hci-kdd.org }
\IEEEauthorblockA{\IEEEauthorrefmark{3}Université de Lorraine \& CNRS, LIEC, UMR 7360, Metz, F-57070 France\\ 
Email: matthieu.geist@univ-lorraine.fr}}


\maketitle

\begin{abstract}
In the era of Internet of Things (IoT) technologies the potential for privacy invasion is becoming a major concern especially in regards to healthcare data and Ambient Assisted Living (AAL) environments.  Systems that offer AAL technologies make extensive use of personal data in order to provide services that are context-aware and personalized. This makes privacy preservation a very important issue especially since the users are not always aware of the privacy risks they could face. A lot of progress has been made in the deep learning field, however, there has been lack of research on privacy preservation of sensitive personal data with the use of deep learning. In this paper we focus on a Long Short Term Memory (LSTM) Encoder-Decoder, which is a principal
component of deep learning, and propose a new encoding technique that allows the creation of different AAL data views, depending on the access level of the end user and the information they require access to. The efficiency and effectiveness of the proposed method are demonstrated with experiments on a simulated AAL dataset.  Qualitatively, we show that the
proposed model learns privacy operations such as disclosure, deletion and generalization and can perform encoding and decoding of the data with almost perfect recovery.
\end{abstract}


% \fi
\IEEEpeerreviewmaketitle

\section{Introduction}
The dramatic demographic change in most western countries will increase the need for development of new Ambient Intelligence (AmI) technologies making use of Artificial Intelligence (AI) and machine learning (ML)\cite{Singh2017}. The new EU Data Protection regulations applying from 2018 onwards \cite{gdpr} will make privacy aware machine learning necessary \cite{PrivacyAwareML}. Consequently, issues of privacy, security, safety and data protection move more and more into the focus of AI and ML, thereby fostering an integrated ML approach \cite{IntegrativeML}, which emphasizes the importance of the human-in-the-loop. Currently, major threats to privacy come from personal data aggregation and the increasing power of data mining and pattern recognition techniques, as well as from healthcare data sharing and analysis. As the number of information sources increases the potential to combine these sources, profile the users and learn sensitive information about them also increases, which makes it a great threat to individual privacy. This an important issue especially in the field of AAL where the users are not always aware of the privacy risks they might face. 

To address the threats mentioned previously we propose a model for the encoding and sharing of combined healthcare and AAL data. The model aims to achieve the privacy of input data before they are distributed to various stakeholders. To protect the privacy a Long Short-Term Memory (LSTM) encoder-decoder system is designed that allows the creation of different data views to correspond to the access level of the receiver.

The remainder of the paper is organized as follows. Section 2 presents an overview of existing privacy techniques and related work for deep learning in privacy. Section 3 introduces the proposed privacy model. Section 4 describes the case study and explains the performed experiments while Section 5 presents the results and the performance of the algorithm. Lastly, Section 6 includes the conclusion and suggestions for future work.

\section{Related Work}
This section is divided into three parts. The first part gives an overview of privacy definitions and identifiers. The second describes previous work done in regards to privacy preserving techniques while the third part gives an introduction to deep learning and overview of existing work in privacy protection with the use of deep learning techniques.

\subsection{Private Data}
The new EU General Data Protection Regulation (GDPR) \cite{gdpr} defines personal data as ``any information relating to an identified or identifiable natural person''  and specifically acknowledges that this includes both `direct' and `indirect' identification. The identification can be by means of ``an identification number or to one or more factors specific to his physical, physiological, mental, economic, cultural or social identity''.
While so far there is not one privacy definition yet that is able to encompass all the different aspects of privacy,  there are guidelines that list the possible identifiers that could be used to identify a person from a group \cite{hipaa} \cite{R97}:

\begin{enumerate}
\item Names, Geographical subdivisions smaller than a state, Dates (other than year)
\item Phone \& Fax Numbers
\item Electronic mail addresses
\item Social Security, Medical Record \&  Health plan beneficiary numbers
\item Account \& Certificate/license numbers
\item Vehicle identifiers and serial numbers (including license plate numbers)
\item Device identifiers and serial numbers
\item Web Uniform Resource Locators (URLs) \& Internet Protocol (IP) address numbers
\item Biometric identifiers, including finger, retinal and voice prints
\item Full face photographic images and any comparable images
\item Any other unique identifying number, characteristic or code
\end{enumerate}

However, the above list of identifiers is not exhaustive, as technology advances more potential identifiers could emerge.

\subsection{Privacy Preservation with Anonymization Methods}

Privacy enhancing technologies protect the users' privacy based on technology, and can offer additional levels of protection than just relying on laws and policies. In order to address the privacy concerns of the users, several approaches have been proposed by the research community. These approaches include information manipulation, privacy and context awareness, access control and data anonymization. In the sections below the anonymization are further analyzed since they are the methods most commonly used for privacy preservation.  
\subsubsection{$k$-Anonymity}
 A very well-known method to anonymize data before releasing them is $k$-anonymity \cite{sweeney2002k}. In a k-anonymized dataset, each record is indistinguishable from at least $k - 1$ other records in regards to specific identifying attributes \cite{xiao2006personalized}. $k$-anonymity is achieved by suppressing (deleting an attribute value from the data and replacing it with a random value that matches any possible attribute value) or generalizing the attributes in the data, which means that an attribute is replaced with a less specific but semantically consistent value \cite{sweeney2002k}. The utility and privacy of the data are connected. There is no way so far that can increase the data privacy without also decreasing the data utility \cite{ohm2009broken}. The objective in these problems is to maximize utility by minimizing the amount of generalization and suppression. Achieving $k$ -anonymity by generalization with this objective as a constraint is a Non-deterministic Polynomial-time hard (NP hard) problem which cannot be solved fully automatically \cite{park2007approximate} \cite{glassboxIML}.  In most cases $k$ -anonymity is able to prevent identity disclosure so that a record in a k-anonymized data set cannot
be connected again to the corresponding record in the
original data set. But in some cases, it may fail to
protect against attribute disclosure. 

\subsubsection{$l$-diversity}

This method was developed to address the weaknesses of $k$-anonymity, which as shown, does not guarantee privacy against adversaries that use background knowledge or in cases where data are lacking diversity. For  $l$-diversity, the anonymization conditions are satisfied if, for each group of records sharing a combination of key attributes, there are at least $l$ -``well-represented" values for each confidential attribute \cite{machanavajjhala2007diversity}. The disadvantage of this method is that it depends on the range of the sensitive attributes. If $l$-diversity is to be applied to a sensitive attribute that does not have many different values, artificial data will have to be inserted. The use of artificial data will improve the privacy but may result in problems with the analysis thus ruining the utility of the data. Also, this method is vulnerable to skewness and similarity attack so it cannot always prevent attribute disclosure. 

\subsubsection{$t$-closeness}
As shown $l$-diversity might not always be sufficient in preventing attribute disclosure. Since it does not account for the semantic closeness of the sensitive values. A  new method named $t$- closeness was proposed in \cite{li2007t} to address these problems. This method requires the distribution of the sensitive attributes in an equivalent class to be close to the distribution of the attribute in the overall table, which in turn means that the distance between the two distributions should be no more than a specified threshold $t$. While the authors in \cite{li2007t} describe ways
to check t-closeness (using several distances between
distributions), no computational procedure to enforce this property is given \cite{domingo2008critique}. The authors proposing the $t$-closing method \cite{li2007t} mention that $t$-closeness limits the amount of useful information that is released. The only way to increase the utility of the data is to increase the threshold t, which in turn decreases the privacy protection.

As seen from the overview of the strengths and weaknesses of each technique $k$-anonymity and the other anonymization methods are not always successful in guarantying that no information is leaked while ensuring usable data levels. While the methods of k-anonymity and $l$-diversity do not always accomplish complete privacy, the method of $t$-closeness provides it. But sometimes it is at the expense of the correlations between confidential attributes and key attributes. Also, the computational method for a specific dataset to be anonymized is an additional problem of this method. The papers defining $k$-anonymity and $l$-diversity propose approaches based on generalization and suppression which a lot of times can cause numerical attributes to become categorical. In the case of $t$-closeness, a computational procedure to reach it is not described. Thus, some issues in this field are still open, both at a conceptual and computational level, which can be improved by defining better properties and by creating more effective methods.

\subsection{Privacy Preservation with Deep Learning}

Deep learning is a promising area of machine learning research with significant success in recent years. So far the applications of deep learning are being used in various systems such as image and speech recognition, data analysis, social media, bioinformatics, medicine, and healthcare. Usually, deep learning architectures are constructed as multi-layer neural networks. There are several different neural network architectures, such as the Recurrent Neural Network (RNN) \cite{goller1996learning}, the feed-forward neural network \cite{bebis1994feed} and the Deep Belief Network (DBN) \cite{hinton2009deep}.
Deep learning has the ability to transform original data into a higher level with more abstract expressions. That means that high-dimensional original data can be converted to low-dimensional data by training  multiple neural networks on how to reconstruct the high-dimensional input data.

However, the existing literature on privacy protection mostly focuses on traditional privacy preserving methods, as described in the previous section, and not on deep learning. Differential privacy proposed by Dwork \cite{dwork2011differential} is one of the few approaches of privacy protection that makes use of machine learning methods. Applications of Differential Privacy include boosting \cite{dwork2010boosting}, principal
component analysis \cite{chaudhuri2013near}, linear and logistic regression \cite{chaudhuri2009privacy, zhang2012functional}
support vector machines \cite{rubinstein2009learning}, risk minimization \cite{chaudhuri2011differentially, wainwright2012privacy} and continuous
data processing \cite{sarwate2013signal}. However, the most relevant work to this paper is that of Dai et al. \cite{dai2013encoder} in which they used an Encoder-Decoder system to protect private information in videos by extracting the privacy region and scrambling it while encoding. The system allows the users to fully restore the original video only if they have a legitimate key, otherwise, they can only see the non private regions in the video.

\section{LSTM Encoder-Decoder Model}

\begin{figure}[h]
\centering
  \includegraphics[] {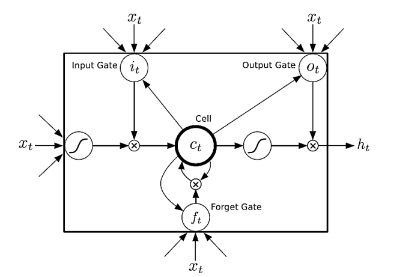}
  \caption{LSTM single cell image~\cite{zhang2015bidirectional}.}
  \label{fig: LSTM}
\end{figure}

Long Short Term Memory networks are a special kind of RNN, capable of learning long-term dependencies \cite{hochreiter1997long}. A basic sequence-to-sequence model, as introduced in \cite{cho2014learning}, consists of two recurrent neural networks (RNNs). The first is an encoder that processes the input and the second a decoder that generates the output. The recurrently connected blocks in LSTM layers are known as memory blocks. Each block contains one or more memory cells which are composed of three units: an input gate, a forget gate and an output gate. These gates modulate the interactions between the memory cell and the environment. Figure \ref{fig: LSTM} shows the single cell of LSTM memory block. LSTM can assist in error minimization because the error can be back-propagated through time and layers. By maintaining a more constant error, the recurrent network can continue to learn over many time steps, and thus be able to link causes and effects. 

The model that was developed in this work uses the multi-layered Long Short-Term Memory (LSTM) encoder to map the input sequence to a vector of a fixed dimensionality, and then another LSTM is used to decode the target sequence from the vector.  The encoder network is the part of the network that takes the input sequence and maps it to an encoded representation of the sequence. The encoded representation is then used by the decoder network to generate an output sequence. This makes the framework have a lock and key analogy, where only someone with the correct key (decoder) will be able to access the resources behind the lock (encoder). The multiple hidden layers of neural networks have characteristics that enable this kind of learning \cite{hinton2006reducing}, along with the mapping characteristic of the encoder-decoder models which are able to create corresponding pairs could make them appropriate for privacy preserving frameworks.

\section{Experiment Design}

The experiments were conducted using as basis the LSTM Encoder-Decoder model introduced in the previous section. In the following sections we describe the evaluation use case scenario and details of the simulated dataset used for the study followed by the modeling and training methodology.

\subsection{Use case}

John is 80 years old and lives in an ambient assisted living environment. He is widowed and was recently diagnosed with Alzheimer's disease. Currently, he lives alone but he likes to stay in touch with his family and friends. The AAL environment he lives in gives him independence and allows him to control his home automation system, for example, he can remotely open and close windows/doors, control the lighting, heating, and the alarm system. Also, it allows monitoring of his vital signs and offers him reminders about medication and appointments. The sensors deployed in the home send the collected information to the cloud offering access to,  family members, care givers, doctors, and researchers.

\begin{figure}[H]
\centering
\includegraphics[width=\linewidth]{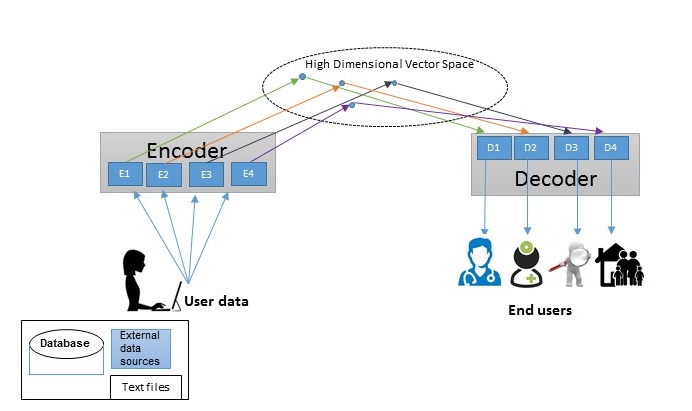}
\caption{Conceptual System Architecture}
\label{fig:architecture2}
\end{figure}

In this use case scenario, four different views of the data are being created  (Figure \ref{fig:architecture2}) depending on the access level of the receiver and the preferences of the user (Table \ref{tab:data}). The user in this scenario has a very close relationship with his family and trusts \cite{TrustInSmartHomes}, \cite{AcceptanceAAL} his caregiver so he has selected almost all his information to be accessible to them, especially because he feels safer knowing they will be able to help him in case of emergency. With regards to doctors, he has allowed only some basic personal and medical information to be visible to them so a different view is created for them. And finally for research purposes, the view that is created does not show any explicit personal data and most of the other sensitive attributes are generalized.

\subsection{Dataset}

For the purposes of this work data related to AAL were simulated. The type of data that were selected are divided into three categories: personal, medical and smart home sensor attributes.
Each entry of the simulated data includes these three kinds of attributes. Personal attributes are those that can explicitly identify a person such as name, address and phone number. The second includes attributes that could potentially identify a person such as gender and birth date. Lastly, the third category includes sensitive medical information and sensor data, like blood pressure, medical history and presence sensors (Table \ref{tab:data}).

\begin{table}[H]
\caption{Simulated Data and Access to Information}
 \label{tab:data}
{\begin{threeparttable}
\begin{tabular}{p{2cm} p{1cm} p{1cm}  p{1cm} p{1cm}}
 \hline

Attribute & Family Member & Doctor & Caregiver & Researcher  \\
 \hline
Name & F & F & F & D \\ 
Age & F & F & G & G \\
Gender & F & F& F& F\\
Height & F & F & G & G \\
Weight & F & F & G & G \\
Address & F & G & F & G \\ 
Phone Number &F & F & F& D  \\
Occupation & F & G & G & G \\
Marital Status  & F & G & G & G\\
Timestamp & F & F & F & F \\
Blood Pressure & G & F & F & G \\
Glucose level & G & F & F & G\\
Disease & F & F& F & G \\
Wearable Pedometer & F & F & F & F\\
Presence Sensor & F & D & F & F\\
Temperature Sensor & F & F & F & G \\
Light Sensor & F & D & D & F \\
Window Sensor & F & D & F & D \\
External Door Sensor & F & D & F & D \\
%Alarm Sensor & F & D & F & D \\
Energy Consumption & G & D& D & G \\ 
\hline
\end{tabular} 
\begin{tablenotes}
      \small
      \item Abbreviations F: fully disclosed, G: generalized, D: deleted
    \end{tablenotes}
  \end{threeparttable}}
\end{table}

The simulation of the data was based on real world data collected from AAL environments and it included personal information, health care data as well as smart home sensor data. The data were simulated for 10000 users, which each user having 100 entries. 

\subsection{Model Configuration}

As described previously, the proposed model makes use of the LSTM neural network
architecture that learns to encode a variable-length input
sequence into a fixed-length vector representation
and to decode a given fixed-length vector representation
back into a variable-length sequence. On this neural network three types of operations are applied to the encoder input by the decoder. These three operations are:  1) Disclosure which means keeping the data as it is 2) Deletion by removing the data or 3) Generalization which means replacing the value with a less specific but semantically consistent value.  So the data from each entry can be fully disclosed to the receiver, generalized or  deleted. Each value for a given attribute has different range aligned with real life values. Four separate views are created for different receivers, family member, doctor, caregiver, and researcher. Each receiver has a different decoder output due to their privacy clearances on patient information (Table \ref{tab:data}).

\begin{figure}[H]
\centering
\includegraphics[width=\linewidth]{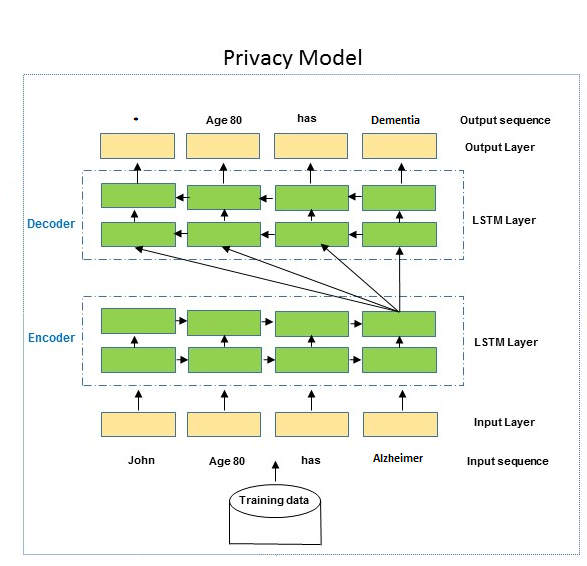}
\caption{Illustration of the proposed LSTM Encoder-Decoder Privacy Model}
\label{fig:model}
\end{figure}

\begin{figure}[H]
\centering
\includegraphics[width=\linewidth]{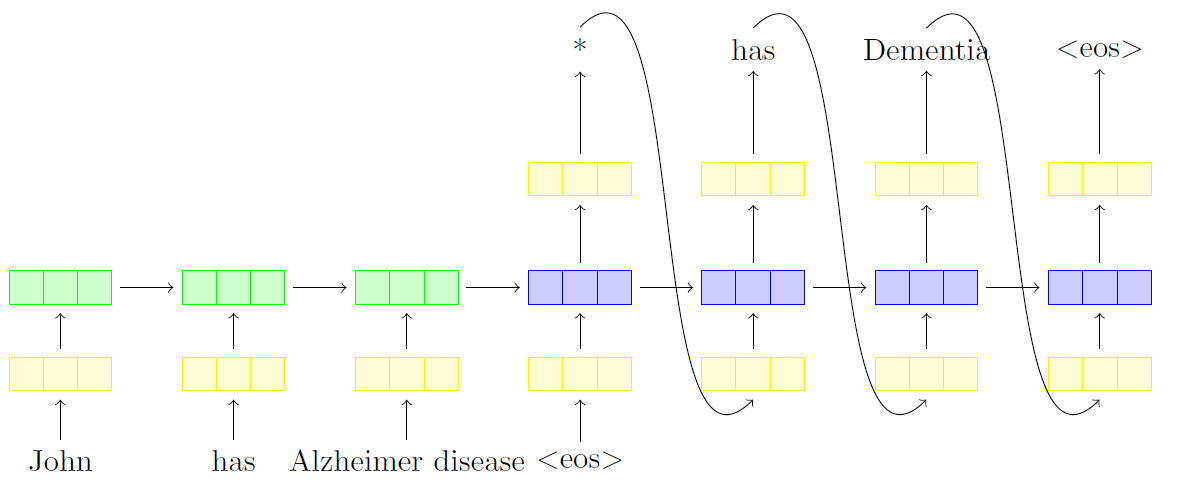}
\caption{Example of the model's Privacy Operations}
\label{fig:example}
\end{figure}

In Figure \ref{fig:model} the overall functionality of the encoder and decoder LSTM layers of the model are depicted. While in Figure \ref{fig:example} we show in more detail how the privacy operations work when the model reads an input sentence such as ``John has Alzheimer'' which will produce ``* has Dementia'' as the output sentence. In this instance the operation of deletion is applied on the Name attribute so `John' is transformed to
`*' and the operation of generalization is applied on the Disease attribute which changes `Alzheimer' to `Dementia'. The model stops making predictions after outputting the end-of-string token `eos'.

\subsection{Model Training and Testing}

For the experiments the models were trained with 800.000 data entries and tested with 200.000 unseen entries. Each user entry consisted of different attribute as shown in Table \ref{tab:data} and their corresponding values, each entry had 160 characters at most. Different attributes are separated with `$|$' in order to easily distinguish between different attributes in an entry.  A 40 character set was used as dictionary. Each sequence is maximum 160 characters long and ends with special token `eos'. In order to handle different sequence lengths, we zero padded each entry to the maximum number of characters which in our case is 160. The encoder and decoder comprise an LSTM network which has 256 hidden units and is trained with Adam \cite{kingma2014adam} with a learning rate of 0.0004.

\section{Results}

We qualitatively analyzed the trained model's results by comparing the decoded outputs
with those of the expected model output after the privacy operations.
The qualitative analysis shows that the LSTM Encoder – Decoder is very good at learning the privacy operations of disclosure, generalization and deletion as well as at capturing the specified preferences in the access to information table. An example of the results of the model can be seen in the following figures.

\begin{figure}[H]
\centering
\includegraphics[width=\linewidth]{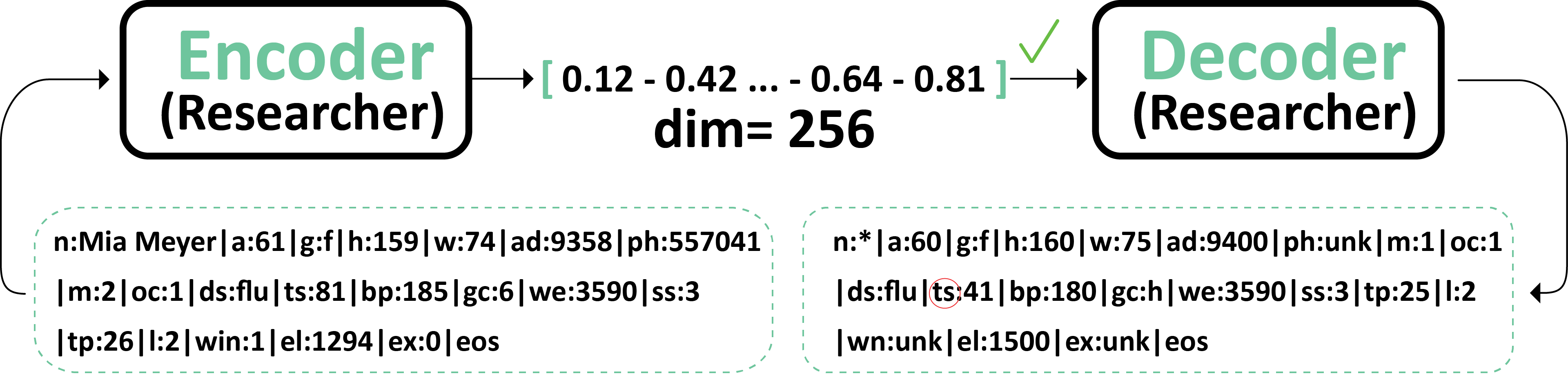}
\caption{Researcher view (With matching Encoder and Decoder)}
\label{fig:Resview}
\end{figure}

 In Figure \ref{fig:Resview}, it can be seen that with the use of the right encoder-decoder mechanism the user's information can be transferred almost perfectly to the researcher with the appropriate privacy rules applied for the researcher's access level. And since only the encoded vector is shared it is not possible to get user information without the right decoder.  

\begin{figure}[H]
\centering
\includegraphics[width=\linewidth]{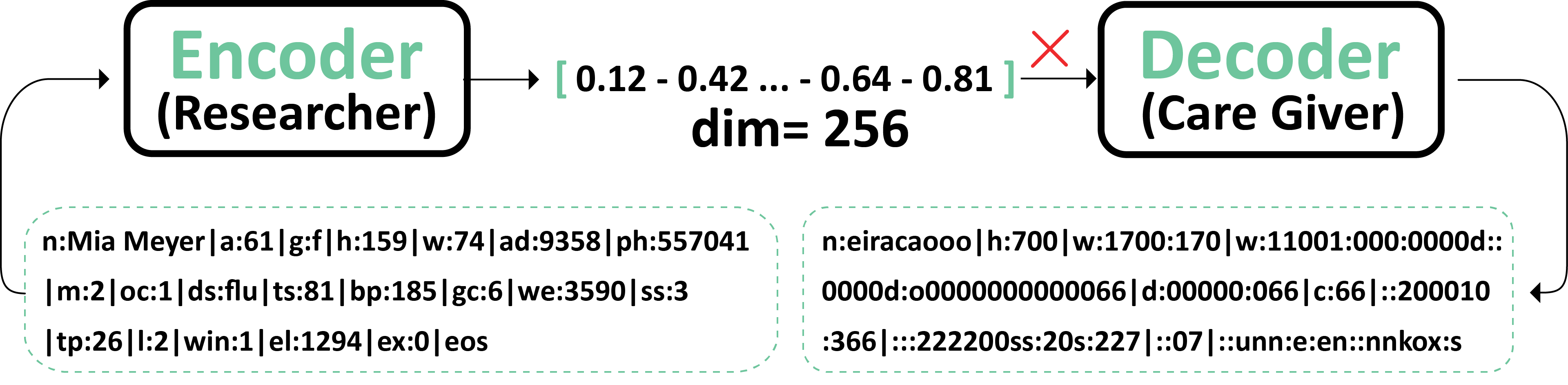}
\caption{Care Giver view (With not matching Encoder and Decoder)}
\label{fig:careview}
\end{figure}

Figure \ref{fig:careview}, showcases one of the most beneficial parts of our model. In this case, we explore the possibility that a Caregiver tries to access the data meant for the Researcher. Because the encoded vector of user information was shared, it would not be possible to decode it unless the right decoder was used. If the encoder and decoder do not match it is not possible to decode and access the user's private information. If the end receiver does not have the correct decoder it is almost impossible to find the right decoder weights since it is very high dimensional floating point vector.

The model has on average 1.5 character error per entry in testing. One or two characters error given the length of 160 characters per entry is very close to perfect recovery. Different decoders are trained for each view and all had the same very close error (1.5 char/entry) during testing. Decoders are shown to be capable of deleting, generalizing or keeping the information given by the encoder. The qualitative analysis of the trained model and the results shows that most of the time the 1 character mistake is in the attribute Timestamp. We attribute this error to the incremental nature of the Timestamp attribute, but this will have to be further investigated with additional experiments. Through these results, we show that the Encoder-Decoder model is able to learn operation rules in a privacy setting by disclosing, generalizing or deleting specific attributes. Thus, this model is able to learn the users' preferences in regards to the privacy policy and create sub-datasets for each receiver with the appropriate information for each one.

\section{Conclusion \& Future Work}
Our LSTM Encoder-Decoder model
is able to learn privacy operations such as disclosure, generalization, and deletion of data, therefore it can generate different data views for data receivers in an encrypted way. It allows users to train independently on their own datasets and selectively share small subsets of their key attributes to specific receivers by creating different views for each one. This offers an attractive point in regards to the utility and privacy trade-off. Two goals are achieved in an end-to-end manner by using the LSTM based encoder and decoder. One is to get an encoded version of user information while the second one is to decode this encoded information according to privacy rules defined by the user. The encoded private information of the user can not be decoded unless the right decoder is used. If an adversary tried using a different decoder on encoded information the system would not disclose any information. Without the right decoder, it would not be possible to train a decoder on encoded information due to the very high dimensionality of the LSTM hidden state vector and possible values of each LSTM network parameter. This way the users preserve the privacy of their data while receiving the benefits of the AAL environment. Moreover, the model can handle raw data even in text format, which is very beneficial in the case of medical records which usually include a lot of doctors' and nurses' notes.  This work is a preliminary experimental study in preserving privacy with the use of LSTM Encoders-Decoders. One of the model's limitations, that will be addressed in future work, is the tokenization of the attributes to improve the performance. Another limitation is the use of simulated data which means they do not contain missing or abnormal values to evaluate the robustness of the method. Future work will also include the expansion of the model to real world data, as well as more complex data formats such as meta-data and multimedia formats.

\section*{Acknowledgments}
This work has been funded by the European Union Horizon 2020 MSCA ITN ACROSSING project (GA no. 616757). The authors would like to thank the members of the project's consortium for their valuable inputs.

%\section*{Acknowledgment}
%This work has been funded by the European Union Horizon2020 MSCA ITN ACROSSING project (GA no. 616757). The authors would like to thank the members of the project's consortium for their valuable inputs.


%\bibliographystyle{IEEEtran}

\bibliography{references}
\bibliographystyle{plain}

\end{document}